%% file: main.tex
\def\review{1} 
\def\arxivdisclaimer{0} 

\documentclass[conference]{IEEEtran}

\usepackage[noadjust]{cite}
\usepackage{amsmath}
\usepackage{amssymb}
\usepackage{amsthm}
\usepackage{amsfonts}
\usepackage{graphicx}
\usepackage{textcomp}
\usepackage{mathtools}
\usepackage{adjustbox}
\usepackage{balance}
\usepackage{bm}
\usepackage{tikz}

\usetikzlibrary{arrows, fit, matrix, positioning, shapes, backgrounds}
\tikzstyle{arrow} = [thick,->,>=stealth]
\usepackage{xcolor}
\def\BibTeX{{\rm B\kern-.05em{\sc i\kern-.025em b}\kern-.08em
    T\kern-.1667em\lower.7ex\hbox{E}\kern-.125emX}}

\usepackage{textcase}
\usepackage[tablename=TABLE,font=small]{caption}
\DeclareCaptionTextFormat{up}{\MakeTextUppercase{#1}}
\usepackage{paralist}
\usepackage{supertabular}    
\usepackage{array} 
\usepackage[linesnumbered,ruled]{algorithm2e}
\usepackage{algorithmic}
\usepackage[acronym]{glossaries}

\if\review1
\usepackage{todonotes}
\else
\usepackage[disable]{todonotes}
\fi

\if\arxivdisclaimer0
\else
\usepackage{hyperref}
\glsdisablehyper
\fi
\usepackage{cleveref}

\usepackage{numprint}     
\usepackage{makecell}   
\usepackage{colortbl}
\usepackage{enumitem}            
\usepackage{longtable} 
\usepackage{wrapfig}
\usepackage{booktabs}
\usepackage{graphics}
\usepackage{pgfplots}
\usepgfplotslibrary{groupplots}
\usepackage{pgfplotstable}
\usepackage{subcaption}
\pgfplotsset{compat=1.18} 

\usepackage{tabu}

\newtheorem*{remark*}{Remark}

\usepackage[utf8]{inputenc}

\crefname{figure}{Fig.}{Fig.}
\crefname{table}{Table}{Table}

\let\oldtabular\tabular
\renewcommand{\tabular}{\small\oldtabular}

\newcolumntype{?}{!{\vrule width 1pt}}
\definecolor{mittelblau}{RGB}{0, 126, 198}
\definecolor{violettblau}{cmyk}{0.9, 0.6, 0, 0}
\definecolor{rot}{RGB}{238, 28 35}
\definecolor{apfelgruen}{RGB}{140, 198, 62}
\definecolor{gelb}{RGB}{1, 221, 0}
\definecolor{orange}{RGB}{244, 111, 33}
\definecolor{pink}{RGB}{237, 0, 140}
\definecolor{lila}{RGB}{128, 10, 145}
\definecolor{hellgrau}{RGB}{224, 224, 224}
\definecolor{mittelgrau}{RGB}{128, 128, 128}
\definecolor{dunkelgrau}{RGB}{80,80,80}
\definecolor{anthrazit}{RGB}{19, 31, 31}
\definecolor{darkgreen}{RGB}{0.125,0.5,0.169}
\definecolor{ahmedyellow}{RGB}{204,153,0}

\begin{document}
%
\title{EMF Exposure Mitigation via \gls{MAC} Scheduling}

\author{
    \IEEEauthorblockN{
    Silvio Mandelli,
        Lorenzo Maggi, 
        Bill Zheng, 
        Azra Zejnilagic, 
        Christophe Grangeat }

	\IEEEauthorblockA{
	Nokia. 
	E-mail: silvio.mandelli@nokia-bell-labs.com}}


\input{Content/Acronyms}

\maketitle

\input{Content/Abstract}
\input{Content/Introduction}
\input{Content/SystemModel}
\input{Content/PowerAllocation}
\input{Content/Simulations}
\input{Content/Conclusion}
\balance
\input{Content/Acknowledgements}

\bibliographystyle{IEEEtran}
\bibliography{references}

\end{document}

%% file: Content/Acronyms.tex
\newacronym{3GPP}{3GPP}{3rd Generation Partnership Project}
\newacronym{5G}{5G}{fifth generation}
\newacronym{6G}{6G}{sixth generation}
\newacronym{CQI}{CQI}{Channel Quality Indicator}
\newacronym{DFT}{DFT}{Discrete Fourier Transform}
\newacronym{DL}{DL}{downlink}
\newacronym{EIRP}{EIRP}{Effective Isotropic Radiated Power}
\newacronym{EMF}{EMF}{electromagnetic field}
\newacronym{EPIC}{EPIC}{EIRP and POLITE Inner-loop Control}
\newacronym{FTP2}{FTP2}{File Transfer Protocol 2}
\newacronym{gNB}{gNB}{g-Node B}
\newacronym{ICNIRP}{ICNIRP}{International Commission on Non-Ionizing Radiation Protection}
\newacronym{IRC}{IRC}{Interference Rejection Combining}
\newacronym{MAC}{MAC}{Medium Access Control}
\newacronym{MCS}{MCS}{Modulation and Coding Scheme}
\newacronym{MIMO}{MIMO}{Multiple Inputs Multiple Outputs}
\newacronym{POLITE}{POLITE}{Power Optimization for Lower Interference and Throughput Enhancement}
\newacronym{FD-POLITE}{FD-POLITE}{Frequency Domain POLITE}
\newacronym{OLLA}{OLLA}{Outer Loop Link Adaptation}
\newacronym{PL}{PL}{Power Limiting}
\newacronym{PRB}{PRB}{Physical Resource Block}
\newacronym{RL}{RL}{Resource Limiting}
\newacronym{SINR}{SINR}{Signal to Interference plus Noise Ratio}
\newacronym{SRS}{SRS}{Sounding Reference Signal}
\newacronym{TDD}{TDD}{Time Division Duplex}
\newacronym{UE}{UE}{User Equipment}
\newacronym{UL}{UL}{uplink}
\newacronym{UMa}{UMa}{Urban Macro}

%% file: Content/Abstract.tex
\begin{abstract}
International standards bodies
define \Gls{EMF} emission requirements that can be translated into control of the base station \emph{actual} \gls{EIRP}, i.e., averaged over a sliding time window.
In this work we show how to comply with such requirements by designing a water-filling power allocation method operating at the \gls{MAC} scheduler level. Our method ensures throughput fairness across users while constraining the \gls{EIRP} to a value that is produced by an outer-loop procedure which is not the focus of our paper. The low computational complexity of our technique is appealing given the tight computational requirements of the \gls{MAC} scheduler.

Our proposal is evaluated against the prior art approaches through massive-MIMO system level simulations that include realistic modeling of physical and \gls{MAC} level cellular procedures. We conclude that our proposal effectively mitigates \gls{EMF} exposure with considerably less impact on network performance, making it a standout candidate for 5G and future 6G \gls{MAC} scheduler implementations.

\end{abstract}
\glsresetall


%% file: Content/Introduction.tex

\section{Introduction} \label{sec:Introduction}

\Gls{5G} cellular network deployments are characterized by the use of massive \gls{MIMO} techniques which ensure high user throughout but also increases Radio Frequency (RF) \gls{EMF} exposure. This trend is expected to continue in \gls{6G} extreme-\gls{MIMO} deployments, where hundreds or even thousands antenna elements, with higher beam gains than in \gls{5G}, are envisioned~\cite{wesemann2023energy}.


When assessing a base station compliance with RF EMF exposure limits~\cite{international2020guidelines}, the actual maximum approach described in~\cite{international2022determination} takes into account the \emph{actual} \gls{EIRP}, i.e., averaged over a time window of 6 to 30 minutes. This window plays a significant role in the assessment of EMF exposure from base stations, due to the variability of their EMF emissions in time and in space. 
Channel modelling studies showed that the actual EIRP threshold is typically a factor 0.25 below the configured maximum EIRP~\cite{baracca2018statistical}, but lower values can be foreseen with advanced beamforming algorithms \cite{rybakowski2023impact}.




Imposing real-time constraints on EIRP in cellular networks requires modifying the \gls{MAC} scheduler operations. This is no easy task, due to the scheduler complexity and the tight computation latency constraint. 
Existing approaches~\cite{tornevik2020time,wigren2021coordinated} propose power control procedures limiting the usable bandwidth; \cite{castellanos2016hybrid} optimizes
precoding for MIMO systems under the assumption of full channel knowledge. However, this assumption is unfeasible in reality as it would require heavy inter-working between physical and \gls{MAC} layer. 
Finally, imposing \gls{EIRP} constraints affects network performance when reaching the actual \gls{EIRP} threshold~\cite{baracca2018statistical}, highlighting the importance of designing algorithms that minimize the performance degradation while being implementable in \gls{MAC} layer real-time procedures. 

\begin{figure}[h]
  \centering
\resizebox{0.95\linewidth}{!}{

\begin{tikzpicture}

\def\lw{1.5}

\node at (-1.2, 2.0) [align=center, rotate=90] (OuterInput) {\large Consumed EIRP \\\large $c_{0,s}, \cdots, c_{t-1,s}$};  
\draw (-0.2, -2.0) rectangle (1.3, 6.0)  node[pos=0.5, align=center, rotate=90] (OuterLoop) {\large Outer-Loop EIRP control};
\draw [arrow, line width=\lw] (OuterInput) -- (-0.2, 2.0) ;

\draw (2.0, -2.0) rectangle (10.0, 6.0) node (Inner Loop) {};
\node at (6.0, 5.5) [align=center] {\large Inner Loop EIRP control (run every slot)};

\draw (2.5, -1.0) rectangle (3.5, 4.5) [red] node[pos=0.5, align=center, rotate=90, red] (EirpBudget) {\large Compute segments budget };

\draw (4.3, -1.4) rectangle (7.5, -0.8) node[pos=0.5, align=center] (TdScheduler) {\large UE Selection};
\draw (4.3, -0.2) rectangle (7.5, 2.7) [red] node[pos=0.5, align=center, red] (POLITE) {\large EIRP-aware \\ \large Power,  MCS, \\  \large PRB limitation};
\draw (4.3, 3.3) rectangle (7.5, 5.0) node[pos=0.5, align=center] (FdScheduler) {\small \large UE Resource \\ \large Allocation};

\draw (8.3, -1.0) rectangle (9.5, 4.5) node[pos=0.5, align=center, rotate=90] (EirpBudget) {\large Update consumed budget};

\draw[arrow, line width=\lw] (1.3, 2,0) -- (2.5, 2.0) node[pos=0.35, above] {\large $\gamma_s^t$};
\draw[arrow, line width=\lw] (3.5, 2,0) -- (4.3, 2.0) node[pos=0.5, above] {\large $b_{s}$};
\draw[arrow, line width=\lw] (3.5, -0.9) -- (4.3, -0.9) node[pos=0.5, above] {\large $b_{s}$};
\draw[arrow, line width=\lw] (5.9, -0.8) -- (5.9, -0.2) node[pos=0.5, right] {UE List};
\draw[arrow, line width=\lw] (5.9, 2.7) -- (5.9, 3.3) node[pos=0.5, right] {\large $A^*_u, r_u, P_u$};.

\draw[arrow, line width=\lw] (7.5, 3.9) -- (8.3, 3.9) node[pos=0.5, above] {\large $A_{u}$};
\draw[arrow, line width=\lw] (7.5, 0.1) -- (8.3, 0.1) node[pos=0.5, above] {\large $P_u$};

\draw [-, line width=\lw] (8.9, -1.0) -- (8.9, -1.6);
\draw [-, line width=\lw] (8.9, -1.6) -- (3.0, -1.6);
\draw [arrow, line width=\lw] (3.0, -1.6) -- (3.0, -1.0) node[pos=0.5, right] {\large $c_{s}$};

\draw [arrow, line width=\lw] (9.5, 2.0) -- (11.0, 2.0) node[pos=0.7, above] {\large $c_{s}^t$};

\draw (11.0, -2.0) rectangle (12.5, 6.0) node[pos=0.5, align=center, rotate=90] {\large Period end: \\ \large Evaluate EIRP consumption $c_{s,t}$};

\draw [-, line width=\lw] (12.0, 6.0) -- (12.0, 7.0);
\draw [-, line width=\lw] (12.0, 7.0) -- (-1.2, 7.0);
\draw [arrow, line width=\lw] (-1.2, 7.0) -- (OuterInput);

\end{tikzpicture}

}
  \caption{Block diagram of \gls{EIRP} control operations considered in this work. Slot index $k$ has been omitted in inner-loop operations for readability. The main focus of our work concerns the red blocks, while the other parts are either legacy \gls{MAC} scheduling operations in the inner-loop, or outer-loop operations discussed in~\cite{maggi2024smooth}.
}
  \label{fig:BlockDiagramGeneral}
\end{figure}
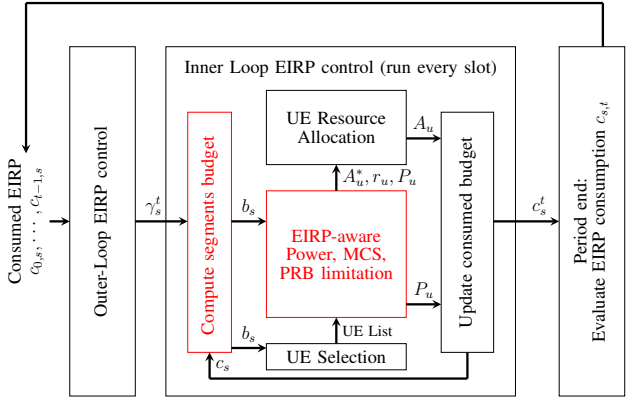

\textbf{Our contribution.} We consider the framework illustrated in Fig.~\ref{fig:BlockDiagramGeneral}, where an ``outer-loop'' mechanism operates on a slow time scale, monitoring the actual \gls{EIRP} over a sliding window of a few minutes, as prescribed in \cite{international2022determination}. Such mechanism sets the \gls{EIRP} budget for the next ``period'' which encompasses the next few hundred of slots. The design of the outer-loop mechanism is not the focus of this paper and can be found, e.g., in~\cite{maggi2024smooth}.
In this paper we focus on the ``inner-loop'' aspect depicted in Fig.~\ref{fig:BlockDiagramGeneral}, which enforces the \gls{EIRP} budget in real-time via \gls{MAC} scheduling operations. Initially, the subset of \glspl{UE} eligible for transmission in the current slot is selected as described, e.g., in~\cite{kela2008dynamic}. 
Then, the \gls{EIRP} is controlled by allocating \glspl{PRB}, power and \gls{MCS} \emph{fairly} across \glspl{UE}.
Our proposal comprises two subroutines: the first spreads the \gls{EIRP} budget across slots within the same period, to accommodate traffic bursts while avoiding budget depletion before the period ends. The second subroutine performs a \emph{fair} allocation across the \glspl{UE} while fulfilling the constraint on the per-slot \gls{EIRP} budget.

Prior art~\cite{tornevik2020time,wigren2021coordinated} proposed constraining resources solely in the frequency, i.e., \gls{PRB}, domain to limit \gls{EIRP}. 
Instead, we propose dynamically adjusting the user \gls{MCS} and power to maximize the global UE throughput fairness while adhering to an \gls{EIRP} constraint. 
The concept of reducing \gls{MCS} and power while spreading the transmission across more radio time and frequency resources was already introduced in~\cite{mandelli2021power,mandelli2022reducing}. Yet, constraints on \gls{EIRP} were not considered since the primary focus was mitigating interference in non-full buffer scenarios.

Our intuition suggests that, to control \gls{EIRP}, reducing power similarly to the approach in~\cite{mandelli2021power,mandelli2022reducing} should be preferred over reducing bandwidth. This reasoning stems directly from the classic Shannon formula: reducing bandwidth leads to a linear decrease in both rate and EIRP, while decreasing power results in a \emph{logarithmic} reduction in rate. These effects tend to coincide in the low SINR regime. 
Therefore, we expect power-based \gls{EIRP} control solutions to drastically outperform bandwidth based approaches, if properly implemented. As a side-product, overall network interference will be reduced, as demonstrated in~\cite{mandelli2021power,mandelli2022reducing}.

We evaluate our proposal in a highly realistic \gls{3GPP} calibrated simulation setup, with a full physical and \gls{MAC} implementation of a massive-\gls{MIMO} cellular scenario. The results enable us to draw conclusions regarding the potential architecture and algorithms for deployment in a \gls{MAC} scheduler of a \gls{gNB} in cellular networks to control \gls{EIRP} with minimal impact on system performance.

%% file: Content/SystemModel.tex
\section{EMF exposure model} \label{sec:SystemModel}

We address the \gls{DL} scheduling problem of allocating \glspl{PRB} and power to users while mitigating the human exposure to EMF according to~\cite{international2022determination}.
We first introduce some notation. We let $A_{u,k}$ and $P_{u,k}$ be the number of allocated \glspl{PRB} and the transmit power per \gls{PRB} for user $u$ in slot $k$, respectively. 
We denote by $\beta_{u,k}$ the beam serving user $u$ in slot $k$ and we call $G_{u,k}(\phi,\theta)$ the antenna gain in azimuth/elevation direction $(\phi,\theta)$ when beam $\beta_{u,k}$ is used. The set of \emph{active} UEs, i.e., the UEs with \gls{DL} data in the buffer, is $\mathcal U_k$.
The EIRP emitted by a radiating antenna array in the azimuth/elevation direction $(\phi, \theta)$ in slot $k$ writes:
\begin{equation}
    \mathrm{EIRP}_{k} (\phi,\theta):=\sum_{u\in\mathcal U_k} A_{u,k} \, P_{u,k} \, G_{u,k}(\phi,\theta).
\end{equation}

The power density measured at distance $R$ from the antenna in the direction $(\phi,\theta)$ is proportional to the EIRP, and equals $\frac{\mathrm{EIRP}(\phi,\theta)}{4\pi R^2}$ in free space conditions. 
When implementing the so-called ``actual maximum approach'' described in \cite{international2022determination}, the \emph{actual} \gls{EIRP}, i.e., time averaged over a sliding window, of the base station shall not exceed a configured actual \gls{EIRP} threshold $\overline{C}$. 
Depending on the environment, different thresholds may need to be configured in different directions. To account for this, we partition the set of all azimuth/elevation angles, defined with respect to the radiating antenna, into a set of \emph{segments} $\mathcal{S}$, and assign a specific actual \gls{EIRP} threshold $\overline{C}_s$ to each segment $s\in\mathcal S$.
We then define the EIRP \emph{consumption} $c_{k,s}$ as the maximum EIRP in segment $s$: 
\begin{align} \label{eq:c_EIRP}
c_{k,s} := & \, \max_{(\phi, \theta)\in s} \mathrm{EIRP}_{k} (\phi,\theta).
\end{align}
We call \emph{period} $t$ a set of $K\ge 1$ consecutive slots starting from $k_0^t$, and we define $c^t_{s}:=\sum_{k=k_0^t}^{k_0^t+K-1} c_{k,s}$ the sum of EIRP consumption over period $t$.
To comply with~\cite{international2022determination}, we impose that the \emph{actual} EIRP, i.e., the average EIRP consumption over every sliding window of $W$ consecutive periods, does not exceed $\overline{C}_s$, for each segment $s$:
\begin{equation} \label{eq:EMFconstr}
	\frac{1}{W} \sum_{i=0}^{W-1} c^{t-i}_s \le \overline{C}_s, \qquad \forall\, t.
\end{equation}

%% file: Content/PowerAllocation.tex
\section{Power control for EMF exposure mitigation} \label{sec:PowerAllocation}

As illustrated in Fig.~\ref{fig:BlockDiagramGeneral}, an outer-loop algorithm computes the limit $\gamma_s^t$ on the EIRP consumption in period $t$, such that
\begin{equation} \label{eq:cap}
0 \le c^t_s \le \gamma^t_s, \qquad \forall\, t, \ s\in \mathcal S.
\end{equation}
The design of the outer-loop algorithm is not the focus of this paper and is discussed, e.g., in~\cite{maggi2024smooth}. Then, we consider $\gamma^t_s$ as a predefined input for each $s,t$.
In this paper we study how to enforce the EIRP cap \eqref{eq:cap} on a per-slot and on a per-user basis via a so-called \emph{inner-loop} control mechanism within \gls{DL} \gls{MAC} scheduler operations depicted in Fig.~\ref{fig:BlockDiagramGeneral}.

Without loss of generality we consider $k_0^t = 1$. We then split our problem into two sub-problems that we solve separately. 
\begin{enumerate}
    \item \textbf{Per-slot EIRP constraint.} For each segment $s$ and in each period $t$, it must hold that $\sum_{k=1}^K c_{k,s}\le \gamma^t_s$. We smooth out the EIRP consumption over time by computing an upper limit - or \textit{slot budget} - $b_{k,s}$ for $c_{k,s}$, for each slot $k$ and segment $s$.
    
    \item \textbf{Power allocation.} At slot $k$, we allocate power fairly across UEs by ensuring $c_{k,s}\le b_{k,s}$.
\end{enumerate}
For simplicity of notation, we herafter drop the dependence of variables on period $t$. 

\subsection{Per-slot EIRP constraint design} \label{sec:perslotEIRPconstraint}

Let us fix the segment $s$. At the beginning of each slot $k=1,\dots,K$, given the past consumption $c_{1,s},\dots ,c_{k-1,s}$, we must cap the EIRP consumption for the upcoming slot to a value $b_{k,s}$. 
We propose to inject a portion $(1-\epsilon_s)$ of the period's budget $\gamma_s$ at slot $k=0$ and spread the remainder by injecting $\epsilon_s\gamma_s/K$ at each slot, with $0\le \epsilon_s \le 1$. Therefore, the available slot budget at slot $k$ is
\begin{equation}
    b_{k,s} = \max \left( \left(1 - \epsilon_s \frac{K-k}{K} \right )\gamma_{s} - \sum_{i=1}^{k-1} c_{i,s}, 0 \right).
    \label{eq:Option1Budget}
\end{equation}
The difficulty in designing the portion $\epsilon_s$ resides in the fact that future user requests are unknown. On the one hand, we want to avoid to greedily deplete the budget early on in the period with $\epsilon_s$ close to zero, which would lead to empty slots later on. On the other hand, we want to refrain from introducing unnecessarily delay by curbing consumption in an overly conservative manner with $\epsilon_s$ close to one.

Similarly, we would like to avoid the possible excessively low slot budget as computed in~\eqref{eq:Option1Budget}. Therefore, the slot budget can be lower bounded by $\rho_s^* c_s^*$, where $c_s^*$ is the maximum \gls{EIRP} that can be radiated in the segment $s$ in one slot on all available \glspl{PRB} and full power and maximum beam gain, and $0 < \rho_s^* c_s^* < \gamma_s/K$.

Optionally, we propose to start curbing \gls{EIRP} once the slot budget is below a certain guard threshold $b_s^*$, controlling \gls{EIRP} emission before the slot budget gets too low. Therefore, we also study the following slot budget refinement  
\begin{equation}
b'_{k,s} =  c_s^* \exp \left( \ln(\rho_s^*) \max( {1 - {b_{k,s}}/{b_s^*}}, 0  )  \right) .
    \label{eq:Option2Budget}
\end{equation}

\subsection{\gls{PL} techniques}

We now show how to enforce the EIRP constraint $b_{k,s}$ on segment $s$ by allocating power to active UEs in a \emph{fair} manner. 
We assume that $A_{u,k}$ PRBs have been pre-allocated to each active UE $u\in\mathcal U_k$ as a function of---amongst others---the amount of bits present in the buffer for each UE, the selected \gls{MCS} and the corresponding rate $m_u, r_u$, respectively.
In our preferred implementation, this is performed via the method described in~\cite{mandelli2022reducing}, which downgrades the \glspl{UE}' \gls{MCS} and reduces the transmit power accordingly, hence increasing the required \gls{PRB} allocation, but without exceeding the \glspl{PRB} available in the current slot. As a result, the transmit power and resulting interference in neighbouring cells are reduced.

\textbf{Two simplifying approximations.} Since power allocation must be decided at every slot, computational complexity is a bottleneck. To simplify the problem we first approximate that each beam $\beta$ has positive gain only in one segment $s(\beta)$, i.e., the one containing its main lobe. 
We call $\mathcal U_{k,s}\subset \mathcal U_k$ the set of active users served by a beam whose main lobe is in segment $s$, i.e., $\mathcal U_{k,s}:=\{u\in \mathcal U_k: \, s(\beta_{u,k})=s\}$. Then, the estimated EIRP consumption $\widehat{c}_{k,s}$ depends only on users $\mathcal U_{k,s}$:
\begin{equation} \label{eq:ref:cks}
    \widehat{c}_{k,s} = \max_{(\phi, \theta)\in s} \sum_{u\in\mathcal U_{k,s}} A_{u,k} \, P_{u,k} \, G_{u,k}(\phi,\theta)
\end{equation}
and the power allocation $P_{u,k}$ for a user in $\mathcal U_{k,s}$ does not impact the EIRP consumption over a different segment $s'\ne s$. In this case, we can optimize the user power allocation for the current slot $k$ independently for each segment.

Next, computing $\widehat{c}_{k,s}$ as in \eqref{eq:ref:cks} requires the solution of an optimization problem. To circumvent it and simplify computations, we pre-compute the maximum gain for each beam:
\begin{equation}
    \widehat{G}_{u,k} := \max_{(\phi, \theta)\in s}  G_{u,k}(\phi,\theta)
\end{equation}
and we bound $\widehat{c}_{k,s}$ via a simple-to-compute expression, that must then be lower or equal than the slot budget:
\begin{equation} \label{eq:bound_c}
\widehat{c}_{k,s} \le \, \sum_{u\in\mathcal U_{k,s}} A_{u,k} \, P_{u,k} \, \widehat{G}_{u,k} \leq b_{k,s}.
\end{equation}
Note that equality in \eqref{eq:bound_c} holds only if the gain is maximal in the same direction for all beams serving users in $\mathcal{U}_{k,s}$. Moreover, the available budget in the current slot is computed in~\eqref{eq:Option1Budget}-\eqref{eq:Option2Budget} with respect to the \emph{actual}, i.e., not approximate, past EIRP consumption values $c_{.,s}$ that are measured in hindsight.

\textbf{Optimization problem.} We now formulate the problem of power allocation across UEs. For notation simplicity we drop the variable dependence on $s,k$, with the understanding that the same procedure is used at each slot $k$ and segment $s$. 

Let $r_{u}(P_u)$ be the function mapping the power allocated to user $u$ to its transmission rate per PRB. 
In practice, $r_{u}$ is a staircase increasing function that can be well approximated as a continuous \emph{concave} function, e.g., $r_{u}(P_{u})=w \log(1+P_{u}/N_{u})$ where $N_{u}$ is the noise plus interference PSD and $w$ is an appropriate constant
for the data channel codes used. 

We allocate power across UEs by maximizing a fairness function of the UE throughput under the EIRP constraint. To this aim, we introduce the fairness function \cite{mo2000fair}:
\begin{equation}
f_\alpha(x):= \left\{ 
	\begin{array}{ll}
		\frac{x^{1-\alpha}}{1-\alpha} \quad \mathrm{if \ } \alpha\ge 0,\ \alpha \ne 1 \\
		\log(x) \quad \mathrm{if \ } \alpha=1
	\end{array} \right. .
\end{equation}
and we formulate our power allocation problem for each segment $s\in \mathcal S$ and slot $k$ as follows:
\begin{subequations}\label{eq:opt}
\begin{align}
    \max_{\{P_{u}\}} & \ \sum_{u \in \mathcal U} f_\alpha\left( A_{u} r_{u}(P_{u})\right) \label{eq:opt1} \\
    \mathrm{s.t.} & \ \sum_{u\in\mathcal U} A_{u} \, P_{u} \, \widehat{G}_{u} \le b \label{eq:opt2} \\
    & \ \underline{P}_{u} \le P_{u} \le \overline{P}_{u}, \quad \forall\, u\in \mathcal U. \label{eq:opt3}
\end{align}
\end{subequations}
where in \eqref{eq:opt2} we used the consumption upper bound \eqref{eq:bound_c}.

Note that, if $\alpha=1$, then the objective function \eqref{eq:opt1} maximizes the proportional fairness of the throughput across UEs; as $\alpha$ grows, it tends to max-min fairness \cite{mo2000fair}. The lower power value $\underline{P}_{u,k}$ can be chosen such that reception is possible with the most robust modulation and coding scheme available.\\

\noindent \textbf{Water-filling solution.} Problem \eqref{eq:opt} is convex, with separable objective function and subject to linear constraints. Hence, if the feasibility region contains an interior point (known as Slater condition \cite{boyd2004convex}), which holds if $\underline{P}_u$'s are sufficiently low, then Karush-Kuhn-Tucker (KKT) conditions are necessary and sufficient conditions for optimality \cite{boyd2004convex}. In our case, KKT conditions lead to the following water-filling type solution.

Define the class of power allocations $\{P_u^*(\nu)\}_u$, depending on the value of a parameter (read, Lagrangian multiplier) $\nu$:
\begin{equation} \label{eq:pnu}
    P_u^*(\nu) \!=\! 
    \left\{
    \begin{array}{ll}
        \!\!\!\!\overline{P}_u, \quad \mathrm{if} \ (A_u \widehat{G}_u)^{-1} \frac{d}{d P_u} f_\alpha(A_u r_u(\overline{P})) > \nu \\
        \!\!\!\!\underline{P}, \quad \mathrm{if} \ (A_u \widehat{G}_u)^{-1} \frac{d}{d P_u} f_\alpha(A_u r_u(\underline{P}_u)) < \nu \\
        \!\!\!\!P^*: \ (A_u \widehat{G}_u)^{-1} \frac{d}{d P_u} f_\alpha(A_u r_u(P^*)) = \nu,  \ \mathrm{else} 
    \end{array}
    \right.
\end{equation}
Then, the solution to the problem \eqref{eq:opt} is $P_u^*=P_u^*(\nu^*)$ for all UEs $u$, where $\nu^*$ is the ``water level'' such that:
\begin{equation} \label{eq:waterlevel}
    \nu^* = \max\left\{ \nu: \ \sum_{u} A_u \widehat{G}_u P_u^*(\nu) \le b \right\}.
\end{equation}
Note that computing $\nu^*$ in \eqref{eq:waterlevel} requires one to find the root of the function $\nu \rightarrow \sum_{u} A_u \widehat{G}_u P_u^*(\nu) - b$, which can be solved numerically via, e.g., the classic bisection method.

%% file: Content/Simulations.tex
\section{Simulations} \label{sec:Simulations}
Our simulation experiments are performed in a \gls{DL} system-level simulator implementing a \gls{3GPP} calibrated \gls{UMa}~\cite{TR38.901} channel model, abstracting the
physical-layer effects through a link to system-level interface computing equivalent \gls{SINR} at transmission time, given the cell/user topology and the beamformed active transmissions in each cell. The simulation environment consists of a hexagonal deployment of seven three-sector sites at $500$~m inter-site distance, corresponding to 21 cells with \gls{gNB} installed at 25~m height and 12 degrees mechanical downtilt. 
The main simulation parameters can be found in Table \ref{tab:SimAssumptions}.
The central frequency is at 3.5 GHz, with 273 \glspl{PRB} at 30 kHz subcarrier spacing generating a 100 MHz carrier in \gls{TDD} split, with an average ratio of 4 \gls{DL} slots every \gls{UL} slot, with a total simulation time of 12 seconds.
In our considered scenario the \glspl{gNB} are equipped with a uniform rectangular array of $12 \times 8$ cross-polarized antennas with 5.2~dBi gain. The \glspl{UE} have 2 cross-polarized omni-directional antenna elements. The antenna spacing is defined in multiple of the wavelength $\lambda$ in Table~\ref{tab:SimAssumptions}.
The wideband \gls{CQI} report is provided every 160 ms to the \gls{gNB}, with measurements taken every 80 ms. \gls{DL} Beamforming is based on Sounding Reference Signals transmitted by the \glspl{UE} in \gls{UL}, with beams selected among a codebook of \gls{DFT} beams~\cite{zoltowski1996closed} in the two-dimensional angular $(\phi, \theta)$ plane.
The \glspl{UE} generate traffic according to the \gls{3GPP} \gls{FTP2} traffic model~\cite{TR36.814}, where each \gls{UE} starts downloading a new packet of $Q$ bits 50 ms after its previous packet was successfully downloaded. 
The \gls{MAC} scheduler architecture is depicted in Fig.~\ref{fig:BlockDiagramGeneral}. At most 8 \glspl{UE} per slot are selected according to proportional-fair metrics in the ``UE selection'' block, and after the possible actions due to actual \gls{EIRP} control, finally the available \glspl{PRB} are allocated to them with a round robin criterion with single-user \gls{MIMO} allocations for up to 4 layers.
In accordance to the proportional-fair criterion, we chose $\alpha = 1$ in~\eqref{eq:opt}.

\begin{table}[ht]%
\centering
\begin{tabular}{|c|c|}
\hline
General Environment & \gls{3GPP} \gls{UMa} \cite{TR38.901}, no buildings
\\ \hline
Cells deployment & $7\times 3$ sector sites at $500$~m distance
\\ \hline
Traffic model & 210 FTP2~\cite{TR36.814}, Reading time $50$ ms
\\ \hline
\gls{CQI} feedback & Wideband reports every 160 ms
\\ \hline
\gls{CQI}/\gls{MCS} Table & CQI/MCS Table 2, 256 QAM~\cite{TS38.214}
\\ \hline
Link Performance & \gls{3GPP} data channel codes~\cite{TS38.214}
\\ \hline
Central Frequency & 3.5 GHz, \gls{TDD}
\\ \hline
Subcarrier Spacing & 30 kHz (0.5 ms slots)
\\ \hline
Number of \glspl{PRB}, band & 273, 100 MHz
\\ \hline
Max Transmit Power, $\overline{P}$ & 53 dBm, $0.73$ W/\gls{PRB}
\\ \hline
\gls{gNB} Antennas, spacing & 12x8 cross-pol, $0.7\lambda$ rows, $0.5\lambda$ cols
\\ \hline
\gls{UE} Antennas, spacing & 2 cross-pol elements $0.5\lambda$
\\ \hline
Spatial multiplexing & Up to rank 4 Single User MIMO
\\ \hline
User Mobility & 3 km/h
\\ \hline
\end{tabular}
\centering
\caption{Main Simulation Parameters}
\label{tab:SimAssumptions}
\end{table}

A single segment is configured to enforce a \gls{EIRP} constraint over the whole sector $\gamma^* = \rho\overline{\gamma}$, where $\overline{\gamma}$ is the maximum \gls{EIRP} that can be radiated by the \gls{gNB} array by using the full 53 dBm transmit power and maximum array gain, and the power reduction factor is $\rho \in (0, 1]$. 
The outer-loop algorithm producing the per-period \gls{EIRP} budget $\gamma$ is implemented according to~\cite{maggi2024smooth}.
The \gls{PL} techniques described in Section~\ref{sec:PowerAllocation} are compared against the ``\gls{RL}'' baseline used in~\cite{tornevik2020time,wigren2021coordinated}, where \glspl{PRB} are consecutively allocated using $P_{u,k}=\overline{P}$ to \glspl{UE} each slot as long as~\eqref{eq:bound_c} is satisfied. 
The slot budget for ``\gls{RL}'' and ``\gls{PL}'' is computed via~\eqref{eq:Option1Budget}.
The ``\gls{PL} - R'' option implements the budget refinement as per~\eqref{eq:Option2Budget}, with $\rho^* = 0.1$ and $b^* = \gamma^*/10$. For reference, the curve without any EIRP control and power optimization~\cite{mandelli2022reducing} is plotted with the label ``No EIRP Control''.

\begin{figure}[t!]
  \centering
  \input{Figures/CellTpFigure}
  \caption{Average cell throughput with different \gls{EIRP} control techniques at constant $\rho=-6$~dB, benchmarked against the achievable performance without any \gls{EIRP} control.} 
  \label{fig:CellTp}
\end{figure}
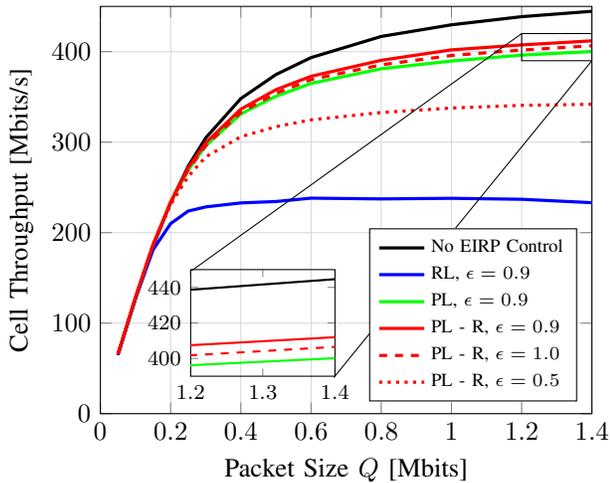

In Fig.~\ref{fig:CellTp} we compare the average cell throughput without actual \gls{EIRP} control against different \gls{EMF}-compliant techniques with $\rho = 1/4$ (or equivalently, $-6$~dB). 
The ``No EIRP Control'' curve shows how the carried load grows with increasing packet size, but saturating after a certain point, reaching 439 Mbits/s (black) at packet size $Q=1.2$~Mbits. 
Once \gls{EIRP} control has been activated, a lower carrier load is to be expected. The \gls{RL} technique (blue) saturates much earlier with a carried load of 237 Mbits/s at $Q=1.2$~Mbits, but still way above $1/4$ of ``No EIRP Control'' performance, thanks to the reduction in interference. Conversely, the \gls{PL} techniques can be much closer to ``No EIRP Control'', achieving up to 407 Mbits/s with the ``\gls{PL} - R'' (red solid) and the hand-tuned $\epsilon = 0.9$. 
One can appreciate the slight yet consistent advantage of ``\gls{PL} - R'' over the simple ``\gls{PL}'' (green), along with the effects of different initial budget allocations in the period, represented by $\epsilon = 1.0, 0.5$ (red dashed, dotted), respectively.

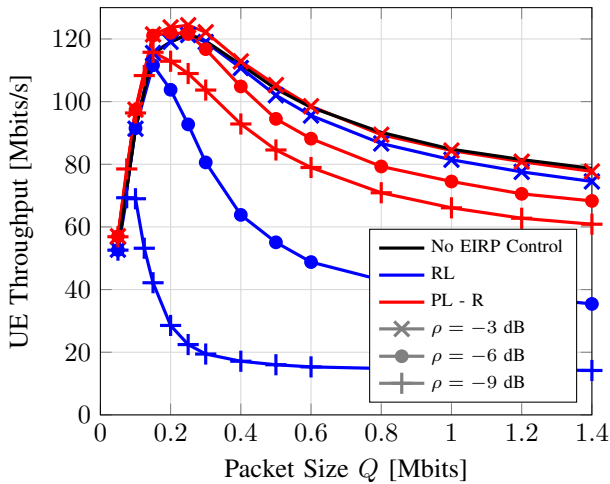
\begin{figure}[t!]
  \centering
  \input{Figures/E2ETpDifferentPrfFigure}
  \caption{Average \gls{UE} throughput (defined as ratio between received bits and the time needed to receive them) without any power adaptation and \gls{EIRP} control, for our best performing ``\gls{PL} - R'' technique, and the baseline ``\gls{RL}'', as the power reduction factor $\rho$ varies.} 
  \label{fig:E2ETpDifferentPrf}
\end{figure}

In Fig.~\ref{fig:E2ETpDifferentPrf} we illustrate the average throughput perceived by each \gls{UE}, defined as the ratio between the bits successfully received and the time taken for their reception. 
After an initial increase in \gls{UE} throughput due to the higher carried traffic in the same time, reductions are observed with large packet size even in the absence of any actual \gls{EIRP} limitation, due to the resource contention among \glspl{UE} and the increased interferece. 
In Fig.~\ref{fig:E2ETpDifferentPrf} we focus on just two \gls{EIRP} control techniques, the best performing ``\gls{PL} - R'' with $\epsilon = 0.9$ (red) versus the \gls{RL} baseline~\cite{tornevik2020time,wigren2021coordinated}. 
Then, we compare their performance with increasing \gls{EIRP} limitations with $\rho=\{-3, -6, -9\}$~dB (cross, circle, plus markers), respectively. As expected, as $\rho$ decreases, the \gls{UE} throughput decreases with smaller packet sizes, corresponding to reduced offered load. However, we can notice the improvements brought by ``\gls{PL} - R'', that are more consistent when the power reduction factor $\rho$ is smaller.
This is due to the superposition of two effects discussed in~\cite{mandelli2021power,mandelli2022reducing}. First of all, the throughput loss scales logarithmically with the transmitted power reduction. Moreover, the power optimization techniques~\cite{mandelli2021power,mandelli2022reducing} applied by \gls{PL}
have the effect of stabilizing and reducing interference, resulting in a potentially higher peak performance. This trend is also evident in Fig.~\ref{fig:E2ETpDifferentPrf}, where we observe gains of 5-8\% at low loads compared to the 'No EIRP Control' case. When $\rho=-3$~dB, the \gls{UE}  throughput experiences a reduction of less than $1\%$ at mid-high loads with ``\gls{PL} - R''.

All proposed \gls{EIRP} control techniques discussed in this work successful enforced the desired EIRP limitation, although we do not provide explicit evidence due to space constraints.



%% file: Figures/CellTpFigure.tex
\tikzset{mark size=2.5}    
    \begin{tikzpicture}
	\begin{axis}[
    		height = 7cm,
    		xlabel={Packet Size $Q$ [Mbits]},
    		ylabel={Cell Throughput [Mbits/s]},
    		ymin=0,
		ymax=450,
		xmin=0,
		xmax=1.4,
		yminorticks = true,
		enlargelimits = false,
		legend pos=south east,
		grid = both,
    		grid style={solid, black!15},
    		legend columns = 1,
   		legend style={font=\scriptsize},
        tick label style={/pgf/number format/fixed},
		legend cell align={left},
		every axis plot/.append style={very thick}
		]

  \addplot[color=black] plot table[x expr=\thisrowno{0}/1.0e3, y expr=\thisrowno{1}/1.0e3, col sep=comma]{Data/CellTpData.csv};
  \addplot[color=blue] plot table[x expr=\thisrowno{0}/1.0e3, y expr=\thisrowno{2}/1.0e3, col sep=comma]{Data/CellTpData.csv};
  \addplot[color=green, solid] plot table[x expr=\thisrowno{0}/1.0e3, y expr=\thisrowno{4}/1.0e3, col sep=comma]{Data/CellTpData.csv};  
  \addplot[color=red, solid] plot table[x expr=\thisrowno{0}/1.0e3, y expr=\thisrowno{5}/1.0e3, col sep=comma]{Data/CellTpData.csv};
  \addplot[color=red, dashed] plot table[x expr=\thisrowno{0}/1.0e3, y expr=\thisrowno{6}/1.0e3, col sep=comma]{Data/CellTpData.csv};
  \addplot[color=red, dotted] plot table[x expr=\thisrowno{0}/1.0e3, y expr=\thisrowno{7}/1.0e3, col sep=comma]{Data/CellTpData.csv};  
		
		

\draw (1.2,390) rectangle (1.4,420);
\draw (1.2,420) -- (0.26,158);
\draw (1.4,390) -- (0.665,40);

	\legend{$\text{No EIRP Control}$, $\text{\gls{RL}, }\epsilon=0.9$, $\text{\gls{PL}, }\epsilon = 0.9$, $\text{\gls{PL} - R, }\epsilon = 0.9$, $\text{\gls{PL} - R, }\epsilon = 1.0$, $\text{\gls{PL} - R, }\epsilon = 0.5$}
   	
   	\end{axis} 

     	\begin{axis}[ymin=390,
		ymax=450,
		xmin=1.2,
		xmax=1.4,
		footnotesize,
		height=3cm, 
		width=3.5cm,
		xtick = {1.2, 1.3, 1.4},
		at={(1.2cm, 0.5cm)},
		every axis plot/.append style={thick} ]

  \addplot[color=black] plot table[x expr=\thisrowno{0}/1.0e3, y expr=\thisrowno{1}/1.0e3, col sep=comma]{Data/CellTpData.csv};  \addplot[color=blue] plot table[x expr=\thisrowno{0}/1.0e3, y expr=\thisrowno{2}/1.0e3, col sep=comma]{Data/CellTpData.csv};
  \addplot[color=green, solid] plot table[x expr=\thisrowno{0}/1.0e3, y expr=\thisrowno{4}/1.0e3, col sep=comma]{Data/CellTpData.csv};  
  \addplot[color=red, solid] plot table[x expr=\thisrowno{0}/1.0e3, y expr=\thisrowno{5}/1.0e3, col sep=comma]{Data/CellTpData.csv};
  \addplot[color=red, dashed] plot table[x expr=\thisrowno{0}/1.0e3, y expr=\thisrowno{6}/1.0e3, col sep=comma]{Data/CellTpData.csv};
  \addplot[color=red, dotted] plot table[x expr=\thisrowno{0}/1.0e3, y expr=\thisrowno{7}/1.0e3, col sep=comma]{Data/CellTpData.csv};  

        \end{axis}

\end{tikzpicture}

%
%
%
%
%
%

%% file: Figures/E2ETpDifferentPrfFigure.tex
\tikzset{mark size=2.5}    
    \begin{tikzpicture}
	\begin{axis}[
    		height = 7cm,
    		xlabel={Packet Size $Q$ [Mbits]},
    		ylabel={\gls{UE} Throughput [Mbits/s]},
        ymin=0,
		ymax=130,
		xmin=0,
		yminorticks = true,
		enlargelimits = false,
		legend pos=south east,
		grid = both,
    		grid style={solid, black!15},
    		legend columns = 1,
   		legend style={font=\scriptsize},
        tick label style={/pgf/number format/fixed},
		legend cell align={left},
		every axis plot/.append style={very thick}
		]

  \addplot[color=blue, solid, mark=x, mark size=4pt, forget plot] plot table[x expr=\thisrowno{0}/1.0e3, y expr=\thisrowno{2}/1.0e3, col sep=comma]{Data/E2ETpDifferentPrfData.csv};
  \addplot[color=blue, solid, mark=*, mark size=2pt, forget plot] plot table[x expr=\thisrowno{0}/1.0e3, y expr=\thisrowno{3}/1.0e3, col sep=comma]{Data/E2ETpDifferentPrfData.csv};
  \addplot[color=blue, solid, mark=+, mark size=4pt, forget plot] plot table[x expr=\thisrowno{0}/1.0e3, y expr=\thisrowno{1}/1.0e3, col sep=comma]{Data/E2ETpDifferentPrf9Data.csv};  
  \addplot[color=red, solid, mark=x, mark size=4pt, forget plot] plot table[x expr=\thisrowno{0}/1.0e3, y expr=\thisrowno{4}/1.0e3, col sep=comma]{Data/E2ETpDifferentPrfData.csv};
  \addplot[color=red, solid, mark=*, mark size=2pt, forget plot] plot table[x expr=\thisrowno{0}/1.0e3, y expr=\thisrowno{5}/1.0e3, col sep=comma]{Data/E2ETpDifferentPrfData.csv};
  \addplot[color=red, solid, mark=+, mark size=4pt, forget plot] plot table[x expr=\thisrowno{0}/1.0e3, y expr=\thisrowno{2}/1.0e3, col sep=comma]{Data/E2ETpDifferentPrf9Data.csv};
  \addplot[color=black] plot table[x expr=\thisrowno{0}/1.0e3, y expr=\thisrowno{1}/1.0e3, col sep=comma]{Data/E2ETpDifferentPrfData.csv};


\def\legy{-1.0}
		\addplot[color=blue, solid, draw=none] coordinates {(0, \legy)};
		\addplot[color=red, solid, draw=none] coordinates {(0, \legy)};
		\addplot[color=gray, solid, mark=x, mark size=4pt, draw=none] coordinates {(0, \legy)};
		\addplot[color=gray, solid, mark=*, mark size=2pt, draw=none] coordinates {(0, \legy)};
		\addplot[color=gray, solid, mark=+, mark size=4pt, draw=none] coordinates {(0, \legy)};

 	\legend{No EIRP Control, \gls{RL}, \gls{PL} - R, $\rho = -3$ dB, $\rho = -6$ dB, $\rho = -9$ dB} 
   	
   	\end{axis} 

\end{tikzpicture}

%
%
%
%
%
%

%% file: Content/Conclusion.tex
\section{Conclusion}\label{sec:Conclusion}

In this work we propose a set of techniques for \gls{EMF} exposure mitigation that control the base station \gls{EIRP} on a slot-per-slot basis to adhere to a budget defined over multiple slots. This budget is produced by an outer-loop algorithm such as the one in \cite{maggi2024smooth}. The integration of these two techniques enables a base station to comply with the ``actual maximum approach'' defined in~\cite{international2022determination}.
Our solution reduces the transmission power fairly across all \glspl{UE} with relatively low computational complexity. Moreover, it is fully parallelizable across different \emph{segments}, that partition the sector into azimuth/elevation angle clusters on which configurable \gls{EIRP} thresholds.

Our proposal significant alleviates the impact of \gls{EMF} mitigation constraints on user performance compared to legacy approaches that only adjust resource allocation without any power adaptation. For example, when the power reduction factor is set to $1/4$ (or $-6$~dB), our method reduces cell throughput at high loads by a mere $6.5\%$, in contrast to a $44.2\%$ reduction with the legacy approaches. 
The minimal computational overhead and the performance achieved in our detailed massive-\gls{MIMO} simulation study confirm that our technique is a standout candidate for enforcing actual \gls{EIRP} constraints in slot-by-slot operations of \gls{5G} and future \gls{6G} \gls{MAC} schedulers.

%% file: Content/Acknowledgements.tex
\section*{Acknowledgements}
The authors thank Tedros Abdu and Alois Herzog for the discussions and feedback during the evolution of this study.
